\documentclass[conference]{IEEEtran}
    \author{
    \IEEEauthorblockN{Luca Bisti\thanks{This work was conducted when the author L. Bisti was working at the Dip. Ingegneria dell'Informazione, University of Pisa}}
    \IEEEauthorblockA{Fluidmesh Networks Inc.\\
    Boston, MA, United States\\
    luca.bisti@fluidmesh.com
    }
    \and
    \IEEEauthorblockN{Luciano Lenzini, Enzo Mingozzi and Carlo Vallati}
    \IEEEauthorblockA{Dip. Ingegneria dell'Informazione, University of Pisa\\
    Via Diotisalvi 2, I-56122 Pisa, Italy\\
    \{l.lenzini, e.mingozzi, c.vallati\}@iet.unipi.it}
    }

\ifCLASSINFOpdf
   \usepackage[pdftex]{graphicx}
\else
   \usepackage[dvips]{graphicx}
\fi
\usepackage{algorithmic}
\usepackage{url}

\begin{document}

%
\title{Efficient handoff for mass transit connectivity using IEEE 802.11}
\maketitle

\begin{abstract}
In the recent years, WiFi standard has been used to develop different kind of wireless networks due to its flexibility and the availability of cheap off the shelf hardware.
Even if the standard itself lacks mobility support, it has been used in networks with mobile nodes.
When mobility is involved, a fast handoff is of paramount importance, especially with multimedia applications.
The current IEEE 802.11 standard does not provide any specification about how the handoff should take place.

In this paper we propose a handoff scheme optimized for networks providing wireless connection to mass transit vehicles.
The structure of the procedure is redesigned to minimize delay as well as assure reliability.
A reliable handoff triggering mechanism is also designed exploiting the results obtained from a set of preliminary outdoor experiments.

The proposed scheme was implemented and deployed in an experimental testbed, and several indoor tests were run in order to demonstrate its reliability and efficiency.

\end{abstract}


%
\IEEEpeerreviewmaketitle


\section{Introduction}

The use of 802.11 technology has been growing exponentially in the last years due to the low cost of hardware and the lack of need for licensed bands.
One of its last application is in the context of providing Internet connection to vehicles of a mass transportation system. The main goal is to bring Internet connection to the passengers, in addition to deploying additional services which need connectivity, e.g. video surveillance, monitoring or ticketing.
Since originally the WiFi standard was not designed to provide connectivity to mobile nodes, the lack of several features for mobility management raises several issues which must be taken into account in such deployments.

In particular, WiFi does not cope with the problem of handoff which, however, is of paramount importance for mobile users, since it takes care of transferring the established connection from one access point to another over the time.
As a matter of facts, the handoff support in IEEE 802.11 standard \cite{80211} is insufficient: a node switches the connection from one AP to another only after the signal strength of the actual connection is so weak that the association is lost.
This procedure can cause a significant disruption of the service which is unfeasible in case of multimedia applications.

Nevertheless, actual mass transit networks have been deployed using 802.11 standards, i.e. \emph{North County Transit District} \cite{nothCounty} and \emph{Seoul Metropolitan Rapid Transit} \cite{seoulMetropolitan}.
Each solution copes with the issues related to mobility, with ad hoc solutions deployed by the manufacturers of the devices, e.g. \cite{belAirNetworks}.
However the proposed solutions are proprietary and they do not assure the compatibility among products of different brands.

In 2008, the IEEE working group ``k'' has standardized a new amendment to the IEEE 802.11 standard in order to add fast handoff support \cite{80211k}.
However this amendment mainly covers the issues related to security and interoperability and it does not define any guidelines on how the handoff itself should be realized.

In this work we present a handoff procedure specifically designed for networks providing connectivity to mass transit vehicles.
We exploit the structure of the network to design an optimized handoff procedure which minimizes the additional delay.
In particular, a set of preliminary experiments has been run to characterize the channel quality over time for a mobile node connecting to multiple APs.
Upon the results of the preliminary measurements, a triggering procedure is designed to assure the reliability of the connection.
The proposal is implemented on our testbed and its performance is assessed through a set of experiments.

The rest of the article is organized as follows. In Section~\ref{sec:architecture} we illustrate the architecture of our testbed. In Section~\ref{sec:measurements} we present the preliminary measurements on the channel quality. In Section~\ref{sec:handoff} we illustrate our proposal, while in Section~\ref{sec:perfeval} the results of the performance evaluation are presented.
Finally, in Section~\ref{sec:conclusions} we draw the conclusions.

\section{Testbed architecture}\label{sec:architecture}

\subsection{Architecture}
%
The WMN testbed we developed is composed by five wireless mesh devices provided by Fluidmesh Networks \cite{fluidmesh}.
Each node is an embedded system used for algorithm prototyping consisting of an x86-compatible, Pentium class CPU, 128 MB of RAM and two 802.11 a/b/g network cards. The hardware device behind each wireless adapter is an Atheros AR5212, one of the most popular chipset commercially available.
On the software side, we have chosen to base our architecture upon the OpenWrt operating system \cite{openwrt}, a Linux-based distribution widely adopted for embedded devices.
In order to add mesh functionalities to the system, we decided to leverage on an existing implementation. \emph{Roofnet} \cite{roofnet} is a wireless mesh network developed by MIT at Cambridge, Massachusetts, with the aim to test new proposals such as routing protocols and rate adaptation algorithms aimed to improve the efficiency of the network itself. Its modular and open source implementation has made it one of the most studied kind of WMNs.
The core of the Roofnet implementation is the \emph{Click Modular Router}, a software program running at the kernel level which is responsible for overriding the normal network stack provided by the operating system. The behavior of a network node can be easily re-defined and adjusted using Click's object model, where packets flow along a collection of interconnected functional elements forming a graph, each of them performing a basic operation such as header manipulation, classification, scheduling and so on.
Roofnet itself is therefore composed by a set of Click elements, i.e. \emph{SRCR}, the routing module. SRCR (SouRCe Routing) is an ad hoc source-based routing algorithm derived from DSR and it is responsible for selecting the path of each flow in the network minimizing, at the same time, the end-to-end transmission time quantified by the Estimated Transmission Time (ETT) metric.

Building on this well-assessed scenario, we have developed a number of enhancements to extend the basic Roofnet system in order to increase the dependability level of the overall network. Specifically, our primary goal was to reduce the disruption time of the packet delivery service, which can be particularly relevant in application scenarios such as video surveillance, site monitoring and remote control. By default, the WMN routing protocol is quite conservative in terms of reactiveness to drastic variations of the link quality or changes in the network topology, e.g., as a consequence of a node fault. This is a classic behavior in WMNs, as transmission errors occur frequently due to the intrinsic nature of the medium, which marks a difference from wired networks, where permanent link failures can be easily detected and several mechanism exists to compensate the fault at the time scale of a few tenths of milliseconds.
SRCR makes no exception to the rule and in its default setup we measured a routes update time in the order of the minute. While it is possible to tweak the configuration of the protocol parameters to speed up the convergence of the routing process, this approach is generally not recommended to avoid excessive route flapping which leads to unstable performance. Therefore the existing architecture should be extended with additional mechanisms which provide the desired resilience improvements without disrupting the routing stability and performance.

\begin{figure}[tbp]
\centering
\includegraphics[width=6.0cm]{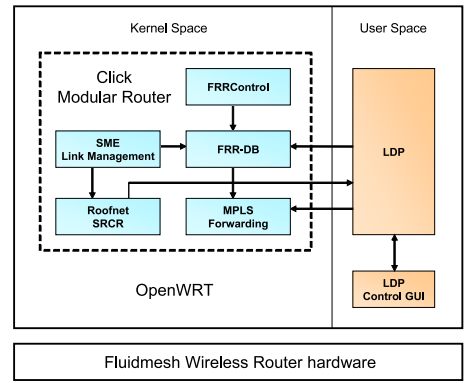}
\caption{Mesh router system architecture.}
\label{fig:routerarch}
\end{figure}
%

Figure \ref{fig:routerarch} illustrates the functional blocks which compose our system and their interactions. As stated before, the OpenWRT operating system and the Roofnet software are the base architecture that provides the foundations to our work. In order to implement the required resilience enhancement that is the objective of our contribution, we have developed a series of modules which will be examined in detail in the rest of this section.

\subsection{MPLS forwarding}
\begin{figure}[tbp]
\centering
\includegraphics[width=4.0cm]{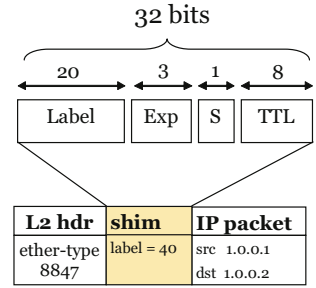}
\caption{MPLS shim header structure.}
\label{fig:shimheader}
\end{figure}

\begin{figure}[tbp]
\centering
\includegraphics[width=6.0cm]{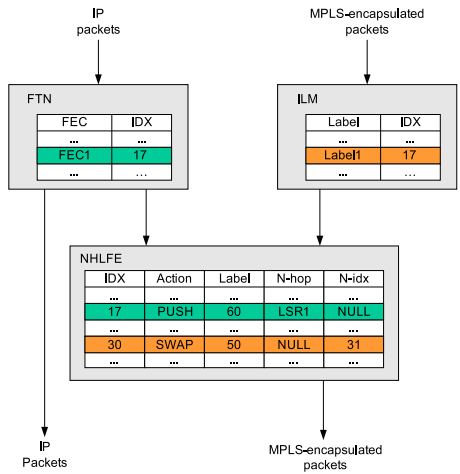}
\caption{MPLS forwarding subsystem.}
\label{fig:implementation}
\end{figure}

The MPLS subsystem implements a fully functional label-switching packet forwarding mechanism operating at an intermediate level between the MAC and the network layers, for which it is often referred to as a layer 2.5 mechanism.
In our WMN architecture it represents a strategic element which provides the foundations to build advanced services such as resiliency enhancements based on fast-rerouting techniques, traffic engineering capabilities, mobility support and so on.
Our design follows the reference MPLS specification \cite{mpls} and supports the standard shim header structure (Figure~\ref{fig:shimheader}), ether-type encapsulation and label stacking capability which should grant for inter-operability with other MPLS implementations.
The implementation consists in three Click router elements which closely match the functional blocks described in the IETF RFC. Figure~\ref{fig:implementation} illustrates the relationship between the modules and their respective data structures.
The \emph{FTN} (FEC-to-NHLFE) element matches ``regular'' ethernet-encapsulated IP packets with a set of configured \emph{Forwarding Equivalence Classes} (FEC), i.e. instances of a packet classification scheme which in the current implementation is represented by the tuple \emph{(src\_ip, src\_port, dst\_ip, dst\_port, proto)}. Whenever a packet is recognized to belong to a known FEC, a 32-bits MPLS shim header is added after its MAC header and then it is emitted from the first output; the label value stored inside the shim header represents an internal index (\emph{IDX}) into the NHLFE table, which will be described later.
Non-matching packets are passed unmodified through the second output for normal processing inside the Roofnet router. The \emph{ILM} (Incoming Label Mapping) element, instead, receives packets already encapsulated with one (or more) MPLS headers as they are being forwarded along an LSP; the top-level label in the stack is simply looked up in the internal table and replaced with the corresponding NHLFE pointer if present, otherwise the whole packet is discarded.
The \emph{NHLFE} (Next-Hop Label Forwarding Entry) element is where the label switching process actually takes place: the internal label value assigned either by the FTN or the ILM elements is used to select a table entry which specifies the operation to be performed on the considered packet.
Implemented operations are: \emph{PUSH}ing a new shim header onto the stack, \emph{POP}ping the top-level header, and \emph{SWAP}ping the label with a different value; the required parameters such as the replacement label value are specified in the various fields of the table.

\section{Preliminary measurements}\label{sec:measurements}
It is well known that the quality of a wireless channel varies almost unpredictably over the time, depending on a number of factors.
In particular, the effects of fading and path loss influence more significantly the channel quality over the time if one of the nodes involved is moving.
We decided to run some preliminary experiments to characterize the channel quality when mobility is considered.
The goal is to monitor the variation of the signal quality over time, without the intervention of any handoff mechanism.
The experience gathered from the results of the experiments is afterwards used to design the handoff scheme, and in particular the triggering mechanism.

In order to have an initial insight on mobility, we preliminary run some indoor tests to collect statistics in a controlled environment.
We setup the experiments in the corridor of our department where the factors which can affect the results are known a priori.
Later on, we run some outdoor scenarios to gather results also with a setup closer to a real scenario.
In the latter we decided to place the experiment in a real urban area, installing one node on the roof of a car moving through the traffic.

We used the same setup for all the preliminary tests conducted, the main difference among them being the location environment and consequently the velocity of the mobile node: pedestrian for indoor, i.e. in the range 3-10 Km/h, and vehicular for outdoor (20-50 Km/h).
Each run consists in a mobile router moving along a pre-established path, with a number of fixed nodes placed aside the route to provide backbone connectivity.
The distance among the fixed nodes is chosen in order to simulate a real handoff scenario, i.e. the two nodes are placed to have their coverage range only partially overlapping at the border.

Table~\ref{table:prelimParams} illustrates the settings.
As it can be seen, the mobile node sends a constant-rate ICMP traffic to all the three fixed routers placed on the path.
This probing traffic enables to collect statistics on both the uplink and downlink directions.
As far as the routing is concerned, a fixed forwarding configuration has been setup using static LSPs on all the nodes, in order to maintain precise control over the routing and to avoid, for instance, undesired flapping  of routes during the experiment.
\begin{table}[]
\centering
\caption{Preliminary experiments settings}\label{table:prelimParams}
\begin{tabular}{|l|l|}
\hline
\bf Name& \bf Value\\
\hline
Number of fixed nodes & 3\\
\hline
PHY mode & 802.11a \\
\hline
Type of traffic & Bidirectional ICMP\\
\hline
Traffic rate & 6 Mbps\\
\hline
Number of packets per second & 75 packets\\
\hline
Routing & Static forwarding using MPLS paths\\
\hline
\end{tabular}
\end{table}

During the experiments, each router collects per-packet statistics at the MAC level, namely \emph{RSSI} (Received Signal Strength Indication) and \emph{packet loss}.
\emph{RSSI} is generically defined by the standard as a numerical value related to the signal level received by the radio antenna, but no exact relationship with physical power scales is given. In fact, the actual meaning of RSSI samples is specific to hardware implementations, i.e. it is defined by the manufacturer of the radio chipset interface.
In our case, for the adopted Atheros-based cards, the RSSI is defined as the \emph{RSSI = SNR+96} in a scale ranging from 0 to 70, where the SNR is the power expressed in dBm. For incoming packets, the reported RSSI value is sampled during the preamble and the PLCP header, while for transmitted packets it is the level at which the ACK packet has been received.
The amount of lost packets over the time is derived from the transmission status report collected for each packet, together with the number of re-transmissions occurred before the successful reception of the ACK.

\begin{figure}[tbp]
\centering
\includegraphics[width=8.0cm]{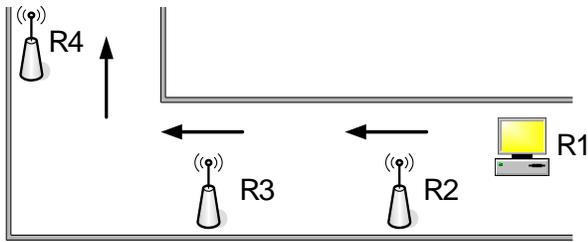}
\caption{Indoor experiment setup.}
\label{fig:indoorPath}
\end{figure}

The configuration is illustrated in Figure~\ref{fig:indoorPath}: three fixed nodes (R2, R3 and R4) are placed along a path traveled by a mobile node (R1).
Each radio interface is configured on the lowest channel of the 802.11a frequency range.
The 802.11a mode is adopted instead of the 'g' in order to avoid interference from the existing wireless networks of the department.
%

\begin{figure}[tbp]
\centering
\includegraphics[width=8.0cm]{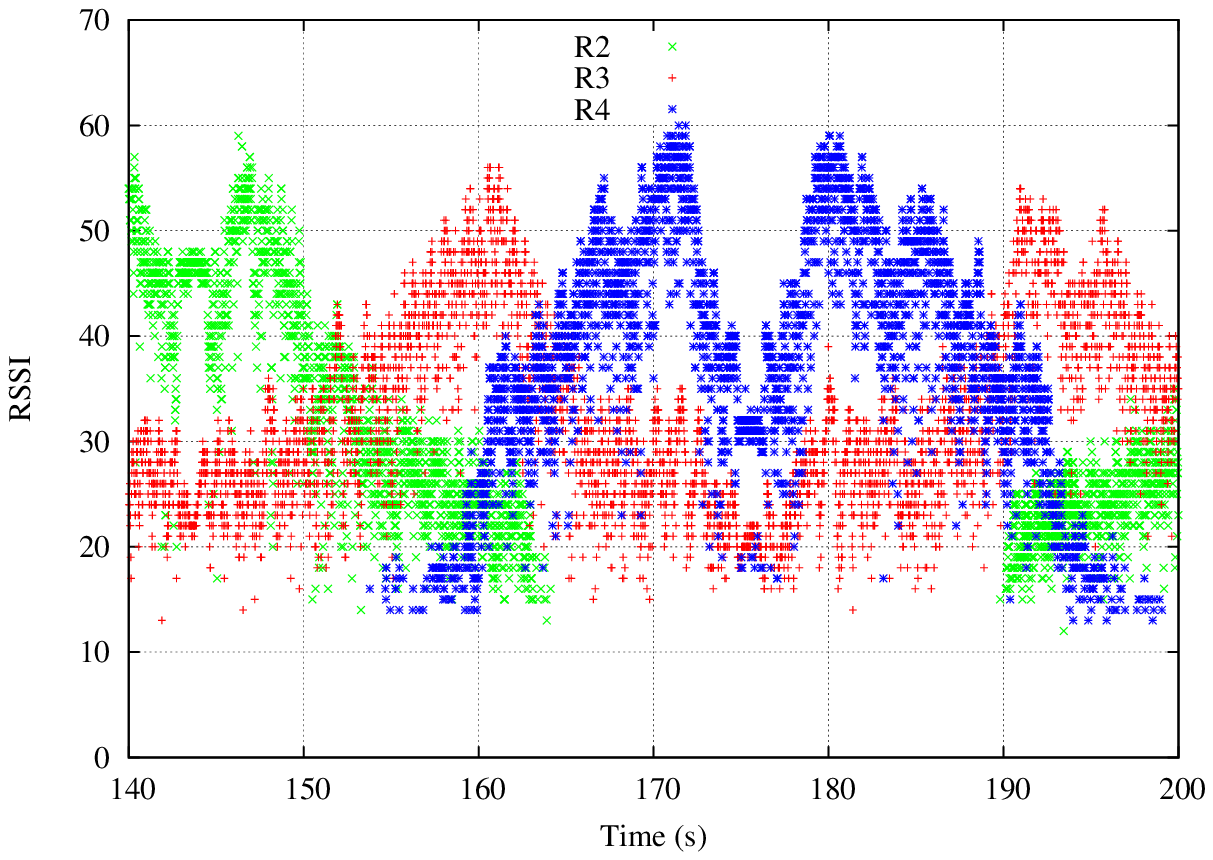}
\includegraphics[width=8.0cm]{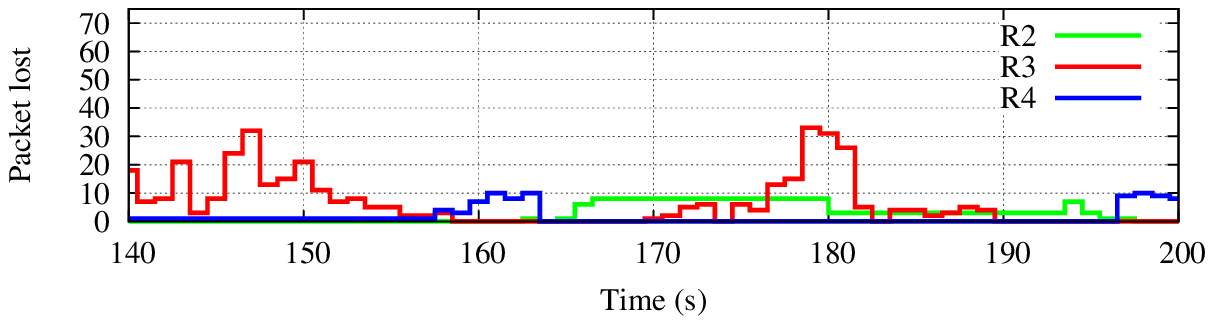}
\caption{RSSI and Packet loss for indoor scenario.}
\label{fig:corridoio}
\end{figure}

Figure~\ref{fig:corridoio} illustrates respectively the RSSI and the packet loss resulting over the time.
For the sake of brevity, we show data relative to the downlink traffic only.
Each curve reports the results related with the traffic from one of the fixed routers.
As already anticipated above, the packet RSSI is measured at the destination when the packet is received by the mobile node, while the packet loss information is collected on the transmitting node.
As can be seen, as the mobile node moves along the path, the transmission quality relative to each fixed router varies according to the distance.
Considering one fixed router at a time, the corresponding RSSI curve assumes a bell shape, and the different bells are shifted over the time by an interval which depends both on the distance between the routers and the absolute speed of the mobile node.
As a matter of fact, the RSSI perceived between two nodes mainly depends on the path-loss which degrades the signal strength proportionally to the distance.
Even if the RSSI has a distinguishable long-term trend, it is characterized by local high-frequency oscillations caused by the channel fading.
Moreover, looking at the same curves, we observe local down-peaks in the middle of the bell where, instead, the highest value is supposed to appear. We believe that this phenomenon is related to the very close distance between the devices antennae, which produces a temporary blinding effect on the radio interfaces.

As far as the packet loss is concerned, we note that strong correlation with the RSSI level can be observed. In particular, when the RSSI is below 26 the packet loss can grow noticeably high, even in the order of 40\%. This kind of simple threshold relationship has been reported to apply for wireless channels operating in low-interference environmental conditions, such as those experienced in rural areas and more in general in absence of  strong RF emissions in the considered spectral bands.
%

Eventually, looking at the results in their entirety, three different zones can be identified in the graph, each one dominated by a particular fixed router. It is straightforward that the handoff mechanism can exploit this peculiarity to maximize the overall quality for packets transmission.

\section{Handoff}\label{sec:handoff}

As previously described, a classic handoff mechanism is composed by three phases, namely \emph{triggering, search} and \emph{execution}.
The overall performance of the scheme is mostly determined by the efficiency of the first two phases: the triggering process is critical as it is responsible for preventing excessive degradation of channel conditions by taking handoff decisions at the most appropriate time, while the search phase causes an interruption of the network service for the time it takes to scan the various channels for another node to connect to; for conventional devices, this time usually amounts to several hundreds of milliseconds, which may cause a significant service disruption.
On the other hand, the execution phase does not introduce a significant delay, since the channel switching operation is usually performed by radio interfaces in a few milliseconds. For this reason, we focus our attention only on the first two phases for our design, assuming a negligible delay for the execution process.

The search phase is the one which can benefit the most from specific optimizations permitted by the particular scenario selected for our study.
Since the path followed by the mobile node is fixed, we can assume that the complete sequence of nodes visited along the path is known a priori. This allows us to scrap the searching phase completely from our design, replacing it with a simple list of node addresses deduced both from the network and the transportation plans.
During the course, a simple table lookup operation is required by the mobile router to obtain the address of the next node to connect to, depending on the current point of attachment to the fixed backbone.
To provide a minimum level of flexibility in the system, the content of the nodes list could be periodically updated by a centralized server in an automatic manner: this would ease the maintenance process of the devices configuration in case of topology changes, as well as fast recovery from possible network failures.

Consequently, the main effort of our contribution has been concentrated on the triggering phase. In order to design an efficient scheme, we strongly leveraged from the experience gained with the results of the preliminary experiments.
As illustrated in Section~\ref{sec:measurements}, RSSI provides a fairly good indication about the quality of the channel and therefore it represents a primary candidate metric for our triggering decision algorithm. For increased reliability in presence of significant radio interference, packet loss information is also considered in some circumstances, as it will be discussed later on.
As shown in our preliminary analysis, raw RSSI samples are affected by short local transients mainly due to the highly time varying channel conditions caused by shadowing and fading.

Therefore, in order to take the correct decisions based on real-time signal strength monitoring, the RSSI values need to be pre-processed to remove high frequency components in the signal. To this aim, after some preliminary evaluations we decided to adopt an exponential weighted moving average (EWMA) filter, which showed to provide adequate signal smoothing characteristics at the price of a modest computational complexity as required for an efficient implementation in real system.
The EWMA filter is defined by the following formula:
\begin{equation}
RSSI_{old}(t) = \alpha RSSI_{curr}(t) + (1-\alpha) RSSI_{old}(t-1)
\end{equation}
where the $\alpha$ parameter is a real number which determines the strength of filtering process. Due to the inability to manipulate floating point values inside the Linux kernel, our EWMA implementation uses a slightly modified version of the formula where $\alpha$ is replaced by the expression $1/2^{s}$, which contains the integer value \emph{s}, called \emph{stability shift} and it can be evaluated without the need of floating point arithmetics.

\begin{figure}[tbp]
\centering
\includegraphics[width=8.0cm]{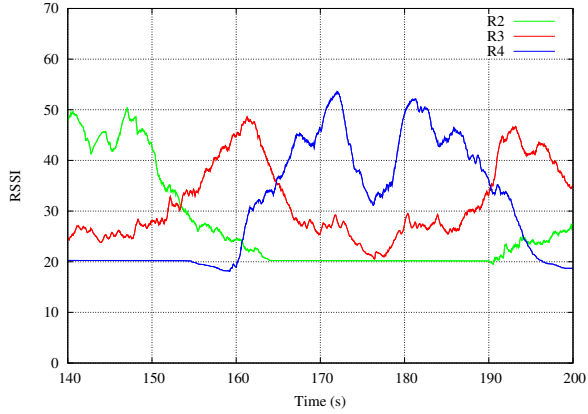}
\caption{Smoothed RSSI curve for indoor scenario (\emph{s=6)}.}
\label{fig:smoothed}
\end{figure}

Figure~\ref{fig:smoothed} shows the same set of data displayed in Figure~\ref{fig:corridoio} after being processed by the EWMA filter configured with a stability shift parameter \emph{s=6}.
In the resulting curves the high-frequency noise oscillations have been strongly attenuated, while the signal still closely follows the short-term trend of the original data. The value of the stability shift parameter can be optimally determined for each specific case of deployment of the system as a trade-off between the desired strength of the low-pass effect and the delay introduced by the filter. In particular, the higher the value of $s$, the cleaner and smoother the output signal is, but at the same time the curves will be shifted to the right in a proportional manner, introducing a delay for any subsequent handoff decisions. Consequently, the maximum speed allowed for the mobile node in order to maintain the desired level of performance will be limited accordingly.

Once that a reliable metric to measure the channel quality was identified, a triggering algorithm has been defined to determine the most appropriate points in time to perform the handoff process basing on the current status of the RSSI levels, so as to minimize the packet loss.
In our solution, the mobile node continuously monitors the signal strength received from the current point of attachment to the network and from the next node expected along the path. The decisional process consists in comparing the two EMWA-averaged signals against single hysteresis thresholds, i.e. when the value of the next node's RSSI is found to be higher than the current node's value by at least $\lambda$ dBm, the handoff is triggered. Such process is run by the mobile node at regular intervals, e.g. 10 times per second; again, an acceptable compromise between responsiveness of the handoff and system overhead needs to be determined.

In order to increase the accuracy, we decided to adopt two different hysteresis thresholds: one ($\lambda_{G}$) is applied when the handoff is going to take place in a region characterized by high RSSI values, while the other ($\lambda_{B}$) is considered in case of low-quality channel conditions, i.e. when significant packet loss is more likely to occur. In our tests we usually assumed $\lambda_{B}<\lambda_{G}$.
This allows to anticipate the connection to the next node when the active link is in unstable conditions.
To distinguish between the two operating regions, another threshold $\beta$ is compared to the current signal level: $\lambda_{G}$ is used in the algorithm if the RSSI is above $\beta$, otherwise $\lambda_{B}$ is considered. Moreover, in the latter case the handoff is triggered only if the packet loss experienced with the next node is low enough ($<P_{L}$) despite the scarce reception quality.
The rationale behind these assumptions is to avoid delaying the switch from the current node too far in time in presence of bad channel conditions and, conversely, to prevent unnecessarily early handoffs when the signal quality stays good during the movement.
The triggering procedure can be summarized through the following pseudo code:
\label{pseudo:alg}
\begin{algorithmic}
\IF {$RSSI_{curr} \geq \beta$}
	\IF {$RSSI_{next} \geq RSSI_{curr} + \lambda_{G}$}
		\STATE Do Handoff
	\ENDIF
\ELSE
        \IF {$RSSI_{next} \geq RSSI_{curr} + \lambda_{B}$ AND $PacketLoss<P_{L}$}
                \STATE Do Handoff
        \ENDIF
\ENDIF
\end{algorithmic}

Packet statistics, including RSSI measurements, can obviously be collected only in presence of network traffic; in general, higher channel utilization levels translate into finer-grained resolution of the RSSI curves and thus better accuracy of the handoff process. Regular data packets transported by the mesh are normally processed by the monitoring subsystem, however some alternate mechanism must also be taken into account to cope with situations where no or insufficient traffic volumes are being exchanged in the network. In particular, this is very likely the case when the mobile node needs to get measurement data from the next node in the path, since it is currently communicating with the attached node and no assumptions can be made concerning independent packet transmissions made by the next node to other routers.
The mechanism provided in our solution consists in the transmission of unicast ICMP background traffic at very low rates (but sufficient to guarantee the required handoff precision) directed towards the requested static node(s). To limit the possible interference caused by the transmissions, the mechanism is activated only when needed and for the minimum time interval required to collect the necessary RSSI samples.

The handoff mechanism has been implemented in our testbed as a dedicated \emph{Click} kernel element integrated into the existing architecture. The module's functionality include handling the probing traffic, collecting and updating the channel statistics and running the triggering algorithm, as well as enforcing the handoff decisions at the networking layer.

\section{Performance evaluation}\label{sec:perfeval}

In this section we present the results obtained from the evaluation of the proposed system.
We repeated the same experiments illustrated in the preliminary analysis with the handoff mechanism activated on the mobile node.
The setup of the experiments is the same illustrated in Table~\ref{table:prelimParams} and the parameters for the handoff algorithm are summarized in Table~\ref{table:handoffParams}.
\begin{table}[]
\centering
\caption{Handoff settings}\label{table:handoffParams}
\begin{tabular}{|l|l|}
\hline
\bf Name& \bf Value\\
\hline
Stability Shift & 6\\
\hline
$\beta$ & 25\\
\hline
$\lambda_{G}$ & 6\\
\hline
$\lambda_{B}$ & 3\\
\hline
$P_{L}$ & 0.5\\
\hline
Probing traffic period & 250ms\\
\hline
\end{tabular}
\end{table}

\begin{figure}[tbp]
\centering
\includegraphics[width=8.0cm]{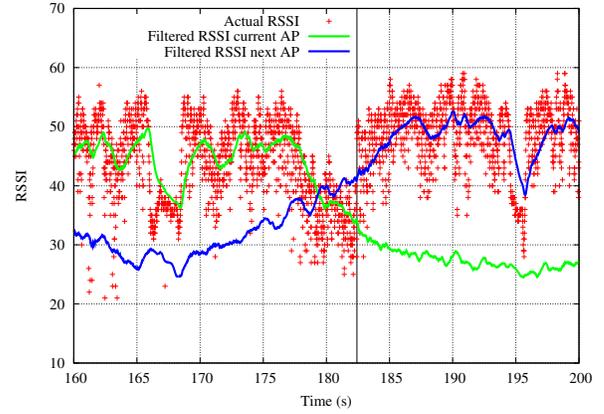}
\caption{Handoff triggering.}
\label{fig:handoff}
\end{figure}

Figure~\ref{fig:handoff} illustrates the obtained results in detail.
The raw values of the actual RSSI experimented by the traffic over a period of 40 seconds are shown in addition to the filtered curves, both for the current and the next nodes.
The RSSI measurements relative to further nodes along the path are not shown since they are not significant for the decisional process performed on the mobile node.

As can be seen, when the RSSI of the current node degrades noticeably, the handoff is triggered at the precise moment indicated by a vertical line.
It is easy to verify that the handoff is activated almost immediately after that the difference between the two RSSI curves becomes greater than $\lambda_{G}$, since the comparison takes place in the "good" region where the signal is above the threshold $\beta$.
Once again, the results demonstrate the effectiveness of the exponential average filter in removing the spurious local oscillations of the RSSI samples.
Besides, looking at the interval between t=160 and t=170, we can see that despite the relevant drop in the current node's RSSI level, the triggering mechanism is correctly prevented from firing.

%

Finally we evaluated the cumulative delay introduced by the handoff process.
The average value measured across the different tests performed was in the order of \emph{20 ms}, which allows for sustained speed of the mobile node.

\section{Conclusions}\label{sec:conclusions}
In this paper we have presented an handoff mechanism for mass transit networks using IEEE 802.11.
The design of our proposal is optimized for this specific case, the goal is to minimize the additional delay.

A set of preliminary experiments has been run to gather information about channel quality when mobility is involved.
The results have been used to design an efficient triggering algorithm for the handoff.
The performance evaluation of our design demonstrates that the handoff delay is minimized.
In addition, the triggering algorithm works well and in particular it helps to switch from one AP to another at the right moment.

\end{document}